\documentstyle[12pt,axodraw,epsfig]{article} 

\newcommand{\be}{\begin{equation}}
\newcommand{\ee}{\end{equation}}
\newcommand{\ba}{\begin{eqnarray}}
\newcommand{\ea}{\end{eqnarray}}

\begin{document}

%\rightline{TIFR/TH/00-51}

\begin{center}
{\large{\bf Jet rates in the hard scattering 
process at finite temperature}}\\
\vspace{1cm}
Krishnendu Mukherjee\footnote{E-mail:km@theory.tifr.res.in}\\
%\vspace{0.5cm}
Department of theoretical Physics,
Tata Institute of Fundamental Research,
Homi Bhabha Road, Mumbai 400 005, India.

\vspace{1cm}

{\bf Abstract}
\end{center}

We compute the cross-section of the hadronic jets arising from the 
quark antiquark pair which are produced from a hard photons 
(of 4-momentum $q$)
in the plasma, predominantly consisting of thermalised quarks and 
gluons. The quark antiquark pair is hard and scattered off the
heat bath to form jets, while the gluons being soft get 
thermalised in the heat bath. The infrared divergences 
cancel in the observable cross-section to $\alpha_s$ order,
which includes the process of emission and absorption of real
gluons. Since the massless quark antiquark pair 
is hard the Compton scattering
processes are absent in the heat bath and it renders an
uncancelled  collinear
divergent piece in the cross-section. We 
regularize it by eliminating the collinear region from the phase space
and write it in terms of jet parameters. The temperature 
dependent part
of the jet cross-section is regular at large ${\sqrt{q^2}\over T}$
and vanishes when ${\sqrt{q^2}\over T}\rightarrow \infty$. 
Since jets carry 
the thermal signature of the hot plasma  
the jet production rate can be used 
as a thermometer of the heat bath.

\vspace{1.5cm}

In this paper we consider a heat bath 
containing plasma of quarks, antiquarks and
gluons at temperature $T$ ($= 1/\beta$). In the $e^+e^-$ collision
a hard photon (hard means it's energy and momentum are much
greater than $T$.) is produced.
Photon enters into the heat bath and produce a hard quark antiquark
pair and a soft gluon (soft means it's energy and  momentum are much
less than $T$.). Since the gluon is soft it gets thermalised, however
the quark antiquark pair being hard, leave the heat bath without
loosing much of their initial energy and form jets.
We have used the real-time formalism of
finite temperature field theory to calculate
the imaginary part of the virtual photon polarization. We take
quarks as massless. The cross-section for the real
processes ($\gamma^*\rightarrow q\bar{q}g$ and 
$\gamma^*g\rightarrow q\bar{q}$) are infrared divergent.
The divergences neatly cancel from the virtual processes
($\gamma^*\rightarrow q\bar{q}$) at $\alpha_s$ order\footnote{At
temperature $T=0$ the KLN \cite{KLN} theorem ensures that
the physically measurable probabilities are infrared finite
even when masses go to zero. Infrared divergences become
severe at $T\neq 0$ than at $T=0$
because the phase space integrals are weighted
by the Bose-Einstein (BE) distribution function. 
In order to get infrared cancellations at $T\neq 0$
KLN prescription is to include 
in the observable rate the emission and absorption
of low energy real particles in the plasma [2-12].}. 
However there still remains an uncancelled part, which
is a collinear divergent piece, appearing from the
real emission process ($\gamma^*\rightarrow q\bar{q}g$).
Since the quark antiquark pair is hard the Compton 
scattering processes ($\gamma^*q\rightarrow\bar{q}g$
and $\gamma^*\bar{q}\rightarrow qg$) are absent in the 
plasma. As a result we are left with this collinear divergent 
piece in the real emission process and hence also in the cross-section.
We have regularised this divergent piece 
by eliminating the collinear region from the phase space 
and write it in terms of the jet parameters.
Since jets carry the information about the plasma 
in the heat bath, the jet production rate (or cross-section)
can be used as a thermometer 
of the heat bath\footnote{Temperature of the plasma 
can also be measured by studying the 
missing-$p_T$ spectrum of multijet events where $p_T$ is the transverse 
component of the total momentum of the jet \cite{Sourendu2}.}  
\cite{Sourendu1}.  

We consider the hard scattering of quark antiquark pair in the
plasma. Since the gluons are very soft they get thermalised in the 
heat bath.
In general, if both quarks and gluons are thermal the imaginary part 
of the photon self energy can be written as \cite{Kobes, Landsman} 
\be
Im\Pi_{\mu\nu}(q) = {i\over 2}{e^{\beta q_0} - 1\over e^{\beta q_0/2}}
\Pi_{\mu\nu}^{12}(q)
\ee
where $\Pi_{\mu\nu}^{12}$ is the $12$ component of the self energy.
In order to obtain the imaginary part for the non-thermal quarks and 
the thermal gluons we will set all the sign function 
in the quark propagators to theta function and also set 
all the fermion distribution functions to zero\footnote{
Massless fermion and gauge field (in 
Feynman gauge) propagators at finite temperature read
$iS_{AB}(p) = p\!\!\!/~ i\tilde{S}_{AB}(p)$ and
$iD^{\mu\nu ab}_{AB}(p) = -\eta^{\mu\nu}\delta^{ab}~ iD_{AB}$ 
($A,B=1,2$) respectively, where $\eta^{\mu\nu}$ is the Minkowski metric 
and $a,b$ ($a,b=1,\cdots,N_c^2-1$ for $SU(N_c)$ gauge group) are
the gauge group indices. $i\tilde{S}_{AB}$ and $iD_{AB}$ are the
$2\times 2$ matrices and their explicit form can be found
in Ref.\cite{Landsman, Niemi}.}. 

The cross-section of $e^+e^-\rightarrow\gamma^*\rightarrow q\bar{q}g$
is given as
\be
\sigma = 2 Im\Pi_{\mu\nu}(q)L^{\mu\nu} .
\ee 
$L^{\mu\nu}$ is the leptonic part and is given as
\be
L^{\mu\nu} = {e^2\over 2(q^2)^3}
(p_1^\mu p_2^\nu + p_1^\nu p_2^\mu - \eta_{\mu\nu} p_1.p_2),
\ee
where $p_1$ and $p_2$ are the 4-momenta of $e^+$ and $e^-$ and $q$
($=p_1+p_2$) is the 4-momentum of the virtual photon. 
The imaginary part of the photon self energy, 
$Im\Pi_{\mu\nu}(q)$, contains the hadronic contributions. 
Since the electromagnetic current is 
conserved  we can write it as 
\be
Im\Pi_{\mu\nu}(q) = - {1\over d-1}\left(-\eta_{\mu\nu} 
+ {q_\mu q_\nu\over q^2}\right) Im\Pi^\mu_{~\mu}(q) 
\ee
where $d$ ($= 4 - 2\epsilon$, $\epsilon>0$) is the space-time dimension.
Since the leptons are taken to be massless, the cross-section reads 
\be
\sigma = - {e^2\over q^4(3-2\epsilon)} Im\Pi^{\mu}_{~\mu}(q) ,
\label{def of cross-section}
\ee
In the following we calculate the scalar $Im\Pi^\mu_{~\mu}$
to obtain the jet cross-section to $\alpha_s$ order.
We shall work here in the rest frame ($\vec{u}=0$) 
of the heat bath.

The loop diagram corresponding to the Born amplitude
of the process $\gamma^*\rightarrow q\bar{q}$ is shown 
in the Fig1. The imaginary part of the trace of 
photon polarisation tensor read 
\ba
Im\Pi^{(B)\mu}_{~~\mu}(q) &=&
(2-d)e^2N_c(\sum_n Q_n^2)q^2\int dR_2 \nonumber\\
&=& - {e^2 N_c\over 4\pi} q^2
\left(q^2\over 4\pi\mu^2\right)^{-\epsilon} 
{\Gamma(2-\epsilon)\over \Gamma(2-2\epsilon)}\sum_n Q_n^2 ,
\ea
where $eQ_n$ is the charge of the $n$-th flavour quark and indices $n$ 
runs over the number of quark flavours. $dR_2$ is the
differential of two particle phase space for massless quarks.
Since the quarks are hard they are non-thermal and the
trace of the polarisation tensor is the same as zero
temperature result.

The two-loop diagrams corresponding to the real 
processes $\gamma^*\rightarrow q\bar{q}g$ 
and $\gamma^*g\rightarrow q\bar{q}$ are shown in the Fig2. 
The complete contribution of these real diagrams to the 
imaginary part of the trace of photon polarisation tensor 
is given as 
\be
Im\Pi^{(R)\mu}_{~~\mu}(q)  = \sum_{j=a,b,c,d}Im\Pi^{(2j)\mu}_{~~~\mu}(q)
= {e^2g^2\over 2}C(R)N_c(\sum_n Q_n^2)\Bigl[ J^{(0)} + J^{(em)}
+ J^{(abs)}\Bigr],
\ee
where
\be
J^{(0)} = \int dR_3 X~ ,~ 
J^{(em)} = \int dR_3 n_B(K_3) X~ ,~
J^{(abs)} = \int d\tilde{R}_3 n_B(K_3) X ,
\label{j-zero, j-em, j-abs}
\ee
and
\be
X(K_1, K_2, K_3) = -{8(1-\epsilon)\over (K_1.K_3)(K_2.K_3)}
[q^2(K_1.K_2) + (K_2.K_3)^2 + (K_1.K_3)^2 - \epsilon(K_1.K_3 + K_2.K_3)^2] .
\label{x mumu}
\ee
Here $C(R)$ is the quadratic Casimir invariant of the representation
$R$ of the quarks. For $SU(N_c)$ gauge group $C(R)=(N_c^2 - 1)/2N_c$,
where the quark furnishes the fundamental representations of the 
group. $K_1$ and $K_2$ are the 4-momenta of the 
quark and antiquark respectively and $K_3$ is
the 4-momentum of the gluon.
$dR_3$ and $d\tilde{R}_3$ are the differential 3-body
phase space for emission and absorption processes 
respectively for massless quarks and gluons\footnote{
3-body phase space for emission process 
has the momentum conserving delta function
$\delta^{(d)}(q - K_1 - K_2 - K_3)$ in $d$ dimension.
To evaluate the phase space integrals for emission 
precess one defines the new variables $x_i = 2q.K_i/q^2$, 
$i=1, 2, 3$ where $x_1 + x_2 + x_3 = 2$ \cite{Muta, Field}.
For absorption process the 3-body phase space integral
contains $\delta^{(d)}(q - K_1 - K_2 + K_3)$.
To evaluate the 3-body phase space integral 
for absorption process one 
goes over to the new variable $x_i$ 
($i=1, 2, 3$) where $x_i$ now
satisfy the relation $x_1 + x_2 - x_3 = 2$.}.

After performing the integrations the final results are 
\ba
J^{(0)}
&=& - {q^2\over 8\pi^3}
\left(q^2\over 4\pi\mu^2\right)^{-\epsilon}
{\Gamma(2-\epsilon)\over \Gamma(2-2\epsilon)}
\biggl[{1\over \epsilon^2} + {1\over \epsilon}\Bigl\{{3\over 2}
- \gamma - \ln\left({q^2\over 4\pi\mu^2}\right)\Bigr\}
+ {1\over 2}(2\gamma -3)\ln\left({q^2\over 4\pi\mu^2}\right)\nonumber\\
& &+ {1\over 2}\ln^2\left({q^2\over 4\pi\mu^2}\right)
- {\gamma\over 2} + {\gamma^2\over 2} - {7\pi^2\over 12}
+ {19\over 4}\biggr] ,
\label{j-zero}\\
J^{(em)}
&=& {q^2\over 8\pi^3}
\left(q^2\over 4\pi\mu^2\right)^{-2\epsilon}
{\Gamma(2-\epsilon)\over \Gamma(2-2\epsilon)}
{\Gamma(1-\epsilon)\over \Gamma(1-2\epsilon)}
\biggl[{2\over \epsilon}\int_0^1 
dx_3 x_3^{-1-2\epsilon}(1-x_3)^{1-\epsilon}
n_B(\sqrt{q^2}x_3/2)\nonumber\\ 
& &+ {1\over \epsilon}{1-\epsilon\over 1-2\epsilon}
\int_0^1 dx_3 x_3^{1-2\epsilon}(1-x_3)^{-\epsilon}n_B(\sqrt{q^2}x_3/2)
- \int_0^1 dx_3 x_3 n_B(\sqrt{q^2}x_3/2)\biggr] ,
\label{j-em}\\
J^{(abs)}
&=& {q^2\over 8\pi^3}
\left(q^2\over 4\pi\mu^2\right)^{-2\epsilon}
{\Gamma(2-\epsilon)\over \Gamma(2-2\epsilon)}
{\Gamma(1-\epsilon)\over \Gamma(1-2\epsilon)}
\biggl[{2\over \epsilon}
\int_0^\infty dx_3 x_3^{-1-2\epsilon}(1-x_3)^{1-\epsilon}
n_B(\sqrt{q^2}x_3/2) \nonumber\\
& &+ {1\over \epsilon}{1-\epsilon\over 1-2\epsilon}
\int_0^\infty dx_3 x_3^{1-2\epsilon}(1-x_3)^{-\epsilon}n_B(\sqrt{q^2}x_3/2)
- \int_0^\infty dx_3 x_3 n_B(\sqrt{q^2}x_3/2)\biggr] .
\label{j-abs}
\ea

Virtual diagrams are shown in the Fig.3 and Fig.4. First we consider
the vertex correction diagrams shown in the Fig.3. 
The complete contribution of the vertex correction diagrams to the
imaginary part of the trace of photon 
polarisation tensor is given as 
\ba
Im\Pi^{(Ve)\mu}_{~~\mu}(q) &=& Im\Pi^{(3a)\mu}_{~~\mu}
+ Im\Pi^{(3b)\mu}_{~~\mu}\nonumber\\
&=& Im\Pi^{(Ve_0)\mu}_{~~~\mu}(q)
+ Im\Pi^{(Ve_T)\mu}_{~~~\mu}(q) .
\ea
Here the $Im\Pi^{(Ve)\mu}_\mu$ is written as the sum of the two 
terms: Zero temperature part and the finite temperature part.
The expression for them are the following:

\ba
Im\Pi^{(Ve_0)\mu}_{~~~\mu}(q)
&=& {e^2g^2\over 16\pi^3}C(R)N_c(\sum_n Q_n^2) q^2
\left(q^2\over 4\pi\mu^2\right)^{-\epsilon}
{\Gamma(2-\epsilon)\over \Gamma(2-2\epsilon)}
\Bigl[{1\over \epsilon^2} + {1\over \epsilon}\Bigl\{{3\over 2}
- \gamma - \ln\left({q^2\over 4\pi\mu^2}\right)\Bigr\}\nonumber\\
& &+ {1\over 2}(2\gamma -3)\ln\left({q^2\over 4\pi\mu^2}\right)
+ {1\over 2}\ln^2\left({q^2\over 4\pi\mu^2}\right)
- {3\over 2}\gamma + {\gamma^2\over 2} - {\pi^2\over 12}
+ 4 \Bigr] ,
\ea
\ba
Im\Pi^{(Ve_T)\mu}_{~~~\mu}(q)
&=& - {e^2g^2\over 16\pi^3}C(R)N_c(\sum_n Q_n^2) q^2
\left(q^2\over 4\pi\mu^2\right)^{-2\epsilon}
{\Gamma(2-\epsilon)\over \Gamma(2-2\epsilon)}
{\Gamma(1-\epsilon)\over \Gamma(1-2\epsilon)}\nonumber\\
& &\Bigl[{4\over \epsilon}\int_0^\infty dx_3 x_3^{-1-2\epsilon}
n_B(\sqrt{q^2}x_3/2)
+ 2\int_0^\infty dx_3 x_3 n_B(\sqrt{q^2}x_3/2)\Bigr] .
\ea

Next consider the self energy correction diagrams in Fig.4. 
The complete contribution of the diagrams in the Fig.4 to the 
imaginary part of the trace of photon polarisation 
tensor is given as 
\ba
Im\Pi^{(Se)\mu}_{~~\mu}(q) &=& - {e^2g^2\over 16\pi^3} C(R) N_c
(\sum_n Q_n^2) q^2 \left({q^2\over 4\pi\mu^2}\right)^{-2\epsilon}
{\Gamma^2(2-\epsilon)\over \Gamma^2(2-2\epsilon)}
{2(1-\epsilon)\over \epsilon} \nonumber\\
& & \times \int_0^\infty dx_3 x_3^{1-2\epsilon}n_B(\sqrt{q^2}x_3/2) .
\ea

To obtain the total cross-section for the 
processes $e^+e^-\rightarrow\gamma^*
\rightarrow q\bar{q}g$ and $e^+e^-\rightarrow\gamma^*g
\rightarrow q\bar{q}$ we have to sum the real and the 
virtual process contributions.
Therefore using the formula in eqn.(\ref{def of cross-section}) the
total cross-section to $\alpha_s$ order reads
\ba
\sigma &=& 
- {e^2\over q^4(3-2\epsilon)}\Bigl[Im\Pi^{(B)\mu}_{~~\mu}(q)
+ Im\Pi^{(R)\mu}_{~~\mu}(q) + Im\Pi^{(Ve)\mu}_{~~\mu}(q)
+ Im\Pi^{(Se)\mu}_{~~\mu}(q)\Bigr] \nonumber\\
&=& \sigma_B\Bigl[ 1 + {\alpha_s\over \pi}C(R)\Bigl\{
\gamma - {\pi^2\over 2} + {3\over 4} + \int_0^1 dx_3 x_3 n_B(\sqrt{q^2}x_3/2)
+ 3 \int_0^\infty dx_3 x_3 n_B(\sqrt{q^2}x_3/2)\nonumber\\ 
& & + F \Bigr\}\Bigr] ,
\label{total corss-section}
\ea 
where $\alpha = {e^2\over 4\pi}$, $\alpha_s = {g^2\over 4\pi}$ and 
$\sigma_B$ is called the Born cross-section and is given as
\be
\sigma_B = {4\pi\alpha^2\over q^2} N_c (\sum_n Q_n^2)
\left({q^2\over 4\pi\mu^2}\right)^{-\epsilon}
{\Gamma(2-\epsilon)\over (3-2\epsilon)\Gamma(2-2\epsilon)} .
\label{born-cross-section}
\ee

Here $F$ contains the finite temperature divergent contributions from 
the real and the virtual diagrams. Its complete expression is  
\ba
F &=& \left({q^2\over 4\pi\mu^2}\right)^{-\epsilon}
{\Gamma(1-\epsilon)\over \Gamma(1-2\epsilon)}
\biggl[{4\over \epsilon}\int_0^\infty dx_3 x_3^{-1-2\epsilon}
n_B(\sqrt{q^2}x_3/2) \nonumber\\
& &+ {2\over \epsilon}
\int_0^\infty dx_3 x_3^{1-2\epsilon}n_B(\sqrt{q^2}x_3/2)\nonumber\\
& & - {2\over \epsilon}\int_0^1 
dx_3 x_3^{-1-2\epsilon}
(1-x_3)^{1-\epsilon} n_B(\sqrt{q^2}x_3/2)\nonumber\\
& & - {2\over \epsilon} \int_0^\infty dx_3 x_3^{-1-2\epsilon}
(1+x_3)^{1-\epsilon} n_B(\sqrt{q^2}x_3/2)\nonumber\\
& &- {1-\epsilon\over \epsilon(1-2\epsilon)}
\int_0^1 dx_3 x_3^{1-2\epsilon} (1-x_3)^{-\epsilon} 
n_B(\sqrt{q^2}x_3/2)\nonumber\\
& &- {1-\epsilon\over \epsilon(1-2\epsilon)}
\int_0^\infty dx_3 x_3^{1-2\epsilon} (1+x_3)^{-\epsilon} 
n_B(\sqrt{q^2}x_3/2)
\biggr] .
\label{A-div}
\ea
After the cancellation of infrared divergences 
between the virtual and real processes finite part of $F$ reads
\ba
F &=& {1\over \epsilon}\int_1^\infty {dx_3\over x_3}
((x_3 - 1)^2 + 1) n_B(\sqrt{q^2}x_3/2)
- 2\int_1^\infty {dx_3\over x_3} 
((x_3-1)^2 + 1)\ln{x_3} n_B(\sqrt{q^2}x_3/2)\nonumber\\
& &+ \int_0^1 {dx_3\over x_3} (1+(1-x_3)^2)\ln(1-x_3)
n_B(\sqrt{q^2}x_3/2)\nonumber\\
& &+ \int_0^\infty {dx_3\over x_3}((x_3 + 1)^2 + 1)\ln(1+x_3) 
n_B(\sqrt{q^2}x_3/2)\nonumber\\ 
& &- \biggl(\gamma + \ln\biggl({q^2\over 4\pi\mu^2}\biggr)\biggr)
\int_1^\infty {dx_3\over x_3} ((x_3-1)^2 + 1) n_B(\sqrt{q^2}x_3/2)\nonumber\\
& &- \int_0^1 dx_3 x_3 n_B(\sqrt{q^2}x_3/2) 
- \int_0^\infty dx_3 x_3 n_B(\sqrt{q^2}x_3/2).
\label{A-finite}
\ea
All the terms in eqn.(\ref{A-finite}) are finite except for the 
first one. The integral in the first term is finite, however due to the
${1\over \epsilon}$ factor before it, it becomes divergent 
when $\epsilon\rightarrow 0$. This divergence appears when the
gluon momentum is parallel to quark or antiquark\footnote{This is a
collinear divergent piece which appears due to the absence of Compton
scattering processes in the heat bath}. 
The natural way of regularizing this divergence is to
eliminate the potentially dangerous collinear integral region 
from the phase space. One can eliminate the regions around 
$\theta =0$ ($0\leq\theta\leq\theta_m$) and
$\theta = \pi$ ($\pi-\theta_m\leq\theta\leq\pi$) 
in the angular integral. Then the infrared divergence 
would have manifested itself as singularity in $\theta_m$
for its vanishing limit.
If the resolving power of the detector is less than $\theta_m$
it can not be able to detect the soft gluons as separate entity
from the quarks. So we can relate $\theta_m$ to the maximum power
of the detector to detect the two particles separately, which
are coming towards it with a small angular separation between themselves.
In order to express this ${1\over \epsilon}$ factor in terms of
$\theta_m$ we look back into the evaluation of $J^{(em)}$ and
the relation is
\be
{1\over \epsilon}\leftrightarrow 2\ln(\theta_m/2) .
\ee
Then the cross-section reads
\be
\sigma = \sigma^{(0)} + \sigma^{(T)},
\ee
where $\sigma^{(0)}$ and $\sigma^{(T)}$ are the zero temperature
and finite temperature piece of the total cross-section respectively.
The expression for them are 
\ba
\sigma^{(0)} &=& \sigma_B \biggl[1 + {\alpha_s\over \pi} C(R)
\biggl\{\gamma - {\pi^2\over 2} + {3\over 4}\biggr\}\biggr] ,
\label{total cross-section at zero}\\ 
\sigma^{(T)} &=& \sigma_B{\alpha_s\over \pi}C(R)
\biggl[ {4\over 3}\pi^2 {T^2\over q^2}
- 2\int_1^\infty {dx_3\over x_3} (1+(x_3 - 1)^2)
\ln{x_3} n_B(\sqrt{q^2}x_3/2)\nonumber\\
& &+ \int_0^1 {dx_3\over x_3}(1+(1-x_3)^2)\ln(1-x_3)
n_B(\sqrt{q^2}x_3/2)\nonumber\\
& &+ \int_0^\infty {dx_3\over x_3}(1+(x_3 + 1)^2)
\ln(1+x_3) n_B(\sqrt{q^2}x_3/2)\nonumber\\
& &+ \biggl(2\ln\biggl({\theta_m\over 2}\biggr) - \gamma
- \ln\biggl({q^2\over 4\pi\mu^2}\biggr)\biggr)
\int_1^\infty{dx_3\over x_3}(1+(x_3-1)^2) n_B(\sqrt{q^2}x_3/2)
\biggr] .
\label{total cross-section at T}
\ea 

We are considering here the hardonic jets arising from the
quark-antiquark pair in $e^+e^-$ annihilation. In 
order that the hadronic jets follow from the quark-antiquark
pair, the quark and antiquark are required not to loose too
much energy in their direction by the emission of gluons and
quark pairs. In otherwords most of the annihilation energy is 
deposited along the direction of the quark and antiquark.
%\subsection{\bf 2-Jet cross-section}
For the production of 2-jets the three body phase space integration
in eqn.(\ref{j-zero, j-em, j-abs}) for emission process must be
restricted. 
We define the contribution of the real diagrams to
the imaginary part of the trace of photon polarisation tensor as
\be
Im\Pi^{(R)\mu}_{{\cal R}\mu}(q)
={e^2g^2\over 2}C(R)N_c(\sum_n Q_n^2)[J^{(0)}_{\cal R}
+ J^{(em)}_{\cal R} + J^{(abs)}]
\ee
where
\ba
J^{(0)}_{\cal R} &=& \int_{\cal R}dR_3 X , 
\label{j-zero-r}\\
J^{(em)}_{\cal R} &=& \int_{\cal R}dR_3 n_B(K_3)X ,
\label{j-em-r} 
\ea
$X$ is given by eqn.(\ref{x mumu}) and the integral 
region ${\cal R}$ is specified by the following conditions: 
The emitted gluon is either soft (i.e. $x_3\leq \Delta$) or collinear to 
one of the quarks (i.e. $\theta_{12}$, $\theta_{23} < 2\delta$)
where $x_3$ ($= 2K_3.q/q^2$)
is the fraction of the energy of the virtual photon carried
by the gluon and $\theta_{13}$ ($\theta_{23}$) is the angle
between the gluon and the quark (antiquark). 
Evaluating the integrals we obtain\footnote{
In addition to $x_i$ ($=2q.K_i/q^2$ where $i=1,2,3$) 
define two more
variables \cite{Muta} $\xi_l=(1-\cos{\theta_{l3}})/2$ ($l=1,2$). 
Among these five variables two are independent and 
so we choose $\xi_1$ and $x_3$ as the independent variables. 
Region ${\cal R}$ is specified by $\xi_{max}\leq\xi_1\leq\xi_{min}$
and $0\leq x_3\leq\Delta$, 
where $\xi_{max}={1-\sin^2\delta\over 1-x_3(2-x_3)\sin^2\delta}$
and $\xi_{min}=\sin^2\delta$. 
Now $J^{(0)}_{\cal R}=J^{(0)}-J^{(0)}_{\bar{\cal R}}$, where 
$\bar{\cal R}$ is obtained by eliminating ${\cal R}$ from the 
whole phase space. So $J^{(0)}_{\bar{\cal R}}
= {q^2\over 16(2\pi)^3}\int_\Delta^1 dx_3 x_3(1-x_3)
\int_{\xi_{min}}^{\xi_{max}}d\xi_1{X\over (1-x_3\xi_1)^2}$
where $X$ is written in terms of $x_3$ and $\xi_1$. Evaluation
of $J^{(em)}_{\cal R}$ is also carried out in the same fashion.}
\ba
J^{(0)}_{\cal R}
&=& - {q^2\over 8\pi^3}
\left(q^2\over 4\pi\mu^2\right)^{-\epsilon}
{\Gamma(2-\epsilon)\over \Gamma(2-2\epsilon)}
\biggl[{1\over \epsilon^2} + {1\over \epsilon}\biggl\{{3\over 2}
- \gamma - \ln\left({q^2\over 4\pi\mu^2}\right)\biggr\}
+ {1\over 2}(2\gamma -3)\ln\biggl({q^2\over 4\pi\mu^2}\biggr)\nonumber\\
& &+ {1\over 2}\ln^2\biggl({q^2\over 4\pi\mu^2}\biggr)
- {\gamma\over 2} + {\gamma^2\over 2} - {11\pi^2\over 12}
+ {13\over 2} - 4\ln\delta\ln\Delta - 3\ln\delta\biggr] 
+ 0(\delta, \Delta) ,
\label{j-zero-rf}
\ea
\ba
J^{(em)}_{\cal R}
&=& {q^2\over 8\pi^3}
\left(q^2\over 4\pi\mu^2\right)^{-2\epsilon}
{\Gamma(2-\epsilon)\over \Gamma(2-2\epsilon)}
{\Gamma(1-\epsilon)\over \Gamma(1-2\epsilon)}
\biggl[{2\over \epsilon}
\int_0^1 dx_3 x_3^{-1-2\epsilon}(1-x_3)^{1-\epsilon}
n_B(\sqrt{q^2}x_3/2)\nonumber\\ 
& &+ {1\over \epsilon}{1-\epsilon\over 1-2\epsilon}
\int_0^1 dx_3 x_3^{1-2\epsilon}(1-x_3)^{-\epsilon}n_B(\sqrt{q^2}x_3/2)
- \int_0^1 dx_3 x_3 n_B(\sqrt{q^2}x_3/2)\nonumber\\
& &- \int_\Delta^1 {dx_3\over x_3} n_B(\sqrt{q^2}x_3/2)
\biggl\{{x_3^2\over 1 - x_3\sin^2\delta}(1-(2-x_3)\sin^2\delta)
+ 2 ((1-x_3)^2 + 1)\ln\tan\delta \nonumber\\
& &+ ((1-x_3)^2 + 1)\ln(1-x_3)\biggr\}\biggr] .
\label{j-em-rf}
\ea

The 2-jets cross-section for the processes $e^+e^-\rightarrow\gamma^*
\rightarrow q\bar{q}g$ and $e^+e^-\rightarrow\gamma^*g 
\rightarrow q\bar{q}$ to $\alpha_s$ order reads
\ba
\sigma_{2jet} &=& 
- {e^2\over q^4(3-2\epsilon)}[Im\Pi^{(B)\mu}_{~~\mu}(q)
+ Im\Pi^{(R)\mu}_{{\cal R}\mu}(q) + Im\Pi^{(Ve)\mu}_{~~\mu}(q)
+ Im\Pi^{(Se)\mu}_{~~\mu}(q)]\nonumber\\
&=&\sigma_{2jet}^{(0)} + \sigma_{2jet}^{(T)},
\label{2-jet cross-section}
\ea
where $\sigma_{2jet}^{(0)}$ and $\sigma_{2jet}^{(T)}$ are 
the zero and finite temperature piece respectively. The
expression for them are the following:
\ba
\sigma_{2jet}^{(0)} &=&
\sigma_B\biggl[ 1 + {\alpha_s\over \pi}C(R)\biggl\{
\gamma - {5\pi^2\over 6} + {5\over 2} - 4\ln\delta\ln\Delta 
- 3\ln\delta\biggr\}\biggr] , 
\label{2-jet cross-section at zero}\\
\sigma_{2jet}^{(T)} &=& \sigma_B{\alpha_s\over \pi}C(R)\biggl[
{4\over 3}\pi^2 {T^2\over q^2} + f_2(\sqrt{q^2}/T, \Delta, \delta)\biggr] ,
\label{2-jet cross-section at T}
\ea
where
\ba
f_2(\sqrt{q^2}/T, \Delta, \delta) &=& - 2\int_1^\infty {dx_3\over x_3} 
n_B(\sqrt{q^2}x_3/2)((x_3 - 1)^2 +1)\ln{x_3}\nonumber\\ 
& &+ \int_0^1{dx_3\over x_3}n_B(\sqrt{q^2}x_3/2)
(1+(1-x_3)^2)\ln(1-x_3)\nonumber\\
& &+ \int_0^\infty {dx_3\over x_3}
n_B(\sqrt{q^2}x_3/2)(1+(x_3+1)^2)\ln(1+x_3)\nonumber\\
& &+ \Bigl(2\ln\delta - \gamma 
- \ln\Bigl({q^2\over 4\pi\mu^2}\Bigr)\Bigr)
\int_1^\infty {dx_3\over x_3} n_B(\sqrt{q^2}x_3/2)((x_3-1)^2 + 1)\nonumber\\
& &+ \int_\Delta^1 {dx_3\over x_3} n_B(\sqrt{q^2}x_3/2)
\Bigl\{x_3^2 + 2((1-x_3)^2 + 1)\ln\delta \nonumber\\
& &+ ((1-x_3)^2 + 1)\ln(1-x_3)\Bigr\}
+ 0(\delta, \Delta),
\label{f2T}
\ea 
and $\sigma_B$ is given by the eqn.(\ref{born-cross-section}).
Here without any loss of generality
we have chosen $\theta_m = 2\delta$. In real life
we can relate $\delta$ to the maximum resolving power 
of the detector.
According to eqn.(\ref{2-jet cross-section at zero}) 
and eqn.(\ref{2-jet cross-section at T}) we see that the
order $\alpha_s$ correction to the two jets from the quark pair 
is controllable in size within the frame work of perturbation theory
at finite temperature. Since the virtual photon is hard,
the ratio ${\sqrt{q^2}\over T}$ is very large. 
The function $f_2(\sqrt{q^2}/T)$ is
well behaved for large ${\sqrt{q^2}\over T}$ ratio and it vanishes
for ${\sqrt{q^2}\over T}\rightarrow \infty$.

3-jets cross-section is the complementary of the 2-jets cross-section.
For 3-jets the imaginary part of the trace of photon polarisation 
tensor reads
\be
Im\Pi^{(R)\mu}_{\bar{{\cal R}}\mu}(q)
= {e^2g^2\over 2}C(R)N_c (\sum_n Q_n^2)
[J^{(0)}_{\bar{\cal R}} + J^{(em)}_{\bar{\cal R}}] .
\ee
The 3-jets cross-section for the process 
$e^+e^-\rightarrow\gamma^*\rightarrow q\bar{q}g$ to $\alpha_s$ 
order reads
\ba
\sigma_{3jet} &=& - {e^2\over q^4(3-2\epsilon)}
Im\Pi^{(R)\mu}_{\bar{{\cal R}}\mu}(q)\nonumber\\
&=& \sigma_{3jet}^{(0)} + \sigma_{3jet}^{(T)} ,
\label{3-jet cross-section}
\ea
where $\sigma_{3jet}^{(0)}$ and $\sigma_{3jet}^{(T)}$ are the 
zero and finite temperature piece respectively. The 
expression for them are 
\ba
\sigma_{3jet}^{(0)} &=& \sigma_B {\alpha_s\over \pi} C(R)
\biggl[4\ln\delta\ln\Delta + 3\ln\delta + {\pi^2\over 3}
- {7\over 3} \biggr] + 0(\delta,\Delta),
\label{3-jet cross-section at zero}\\
\sigma_{3jet}^{(T)} &=& \sigma_B {\alpha_s\over \pi}C(R)
f_3(\sqrt{q^2}/T) + 0(\delta, \Delta) ,
\label{3-jet cross-section at T}
\ea
where
\be
f_3(\sqrt{q^2}/T) = - \int_\Delta^1 {dx_3\over x_3}
n_B(\sqrt{q^2}x_3/2)
\Bigl\{x_3^2 + 2 ((1-x_3)^2 + 1)\ln\delta 
+ ((1-x_3)^2 + 1)\ln(1-x_3)\Bigr\} .
\label{f3T}
\ee
Here $f_3(\sqrt{q^2}/T)$ is also well behaved for large 
${\sqrt{q^2}\over T}$ ratio and it vanishes for 
${\sqrt{q^2}\over T}\rightarrow\infty$. 
Because of $\ln\delta$ and $\ln(1-x_3)$ in the 
second and third term  
within the second bracket of eqn.(\ref{f3T}) respectively we obtain 
positive contribution after integrations. However
the first term within the bracket will give 
negative contribution. We write the 3-jets cross-section as
\be
\sigma_{3jets} = \sigma_B {\alpha_s\over \pi} C(R) 
h(\sqrt{q^2}/T, \delta, \Delta) + 0(\delta, \Delta),
\ee
where
\be
h = 4\ln\delta\ln\Delta + 3\ln\delta + {\pi^2\over 3}
- {7\over 3} + f_3(\sqrt{q^2}/T) .
\ee
The plots of $h$ versus $\delta$ (Fig.5 and Fig.6) 
for different set of values of
$\sqrt{q^2}/T$ and $\Delta$ show that the function $h$ remains positive
for realistic values of its argument. The maximum value of $\Delta$
we have taken is $0.4$, which is much away from the soft gluon 
approximation, yet $h$ remains positive for this large value of
$\Delta$. 

We obtain from eqn.(\ref{2-jet cross-section}) and
eqn.(\ref{3-jet cross-section}) that the ratio
of the two cross-section is
\be
{\sigma_{3jet}\over \sigma_{2jet}}
= {\alpha_s\over \pi} C(R) h(\sqrt{q^2}/T, \delta, \Delta) + 0(\alpha_s^2).
\ee
Apart from the small numerical factor 
${\alpha_s\over \pi}C(R)$
this ratio behaves in the same way as the 
function $h$. As $\delta$ and $\Delta$ decrease
the 3-jets phase space gets increased compared
to the 2-jets, as a result this ratio gets 
increased which are evident from the plots
in Fig.5 and Fig.6. It is also evident from these
plots that for a fixed value of $\delta$ and $\Delta$
as $\sqrt{q^2}/T$ increases the phase space of gluons
get decreased due to Bose-Einstein distribution 
function and as a result this ratio gets decreased.

I gratefully acknowledge S. Gupta for suggesting the problem
and for the many subsequent discussions and helpful comments.
I also thank P. Roy and K. Sridhar for helpful discussions and
suggestions. I am also thankful to P. Aurenche for clarifying certain
points regarding the infrared cancellations in their paper \cite{Altherr1}
and also for helpful comments.

\newpage

%%%%%%%%%%%%% Fig.1 %%%%%%%%%%%%%%%%%%%%%%%%
\begin{center}
\begin{picture}(300, 100)(0,0)
%\Line(0,0)(300,0)
%\Line(0,0)(0,100)
\ArrowArcn(150,50)(40,180,0)
\ArrowArcn(150,50)(40,0,180)
\Photon(50,50)(110,50){3}{4}
\Photon(250,50)(190,50){3}{4}
\Text(108,43)[]{1}
\Text(194,44)[]{2}
\Text(150,100)[]{$K_1$}
\Text(150,0)[]{$K_2$}

\end{picture}\\
{Figure 1: {\sl One loop photon polarization graph for the computation of 
Born amplitude.}}

\end{center}

%%%%%%%%%%%%%% End Fig.1 %%%%%%%%%%%%%%%%%%%%%

\vspace{2.5cm}

%%%%%%%%%%%%%%% Fig.2  %%%%%%%%%%%%%%%%%

\begin{center}
\begin{picture}(450,250)(0,0)
%\Line(0,0)(450,0)
%\Line(0,0)(0,250)
%%%%%% (a) %%%%%%%
\ArrowArcn(110,200)(40,180,90)
\ArrowArcn(110,200)(40,90,0)
\ArrowArcn(110,200)(40,0,90)
\ArrowArcn(110,200)(40,90,180)
\Photon(10,200)(70,200){3}{4}
\Photon(210,200)(150,200){3}{4}
\Gluon(110,240)(110,160){3}{7}
\Text(68,193)[]{$1$}
\Text(155,194)[]{$2$}
\Text(110,247)[]{$1$}
\Text(110,153)[]{$2$}
\Text(110,135)[]{$(a)$}
%%%%%% (b) %%%%%%%
\ArrowArcn(340,200)(40,180,90)
\ArrowArcn(340,200)(40,90,0)
\ArrowArcn(340,200)(40,0,90)
\ArrowArcn(340,200)(40,90,180)
\Photon(240,200)(300,200){3}{4}
\Photon(440,200)(380,200){3}{4}
\Gluon(340,240)(340,160){3}{7}
\Text(298,193)[]{$1$}
\Text(386,194)[]{$2$}
\Text(340,247)[]{$2$}
\Text(340,153)[]{$1$}
\Text(340,135)[]{$(b)$}
%%%%%% (c) %%%%%%%%%%
\ArrowArcn(110,80)(40,180,135)
\ArrowArcn(110,80)(40,135,45)
\ArrowArcn(110,80)(40,45,0)
\ArrowArcn(110,80)(40,0,180)
\Photon(10,80)(70,80){3}{4}
\Photon(210,80)(150,80){3}{4}
%\Gluon(110,120)(110,40){3}{7}
\GlueArc(110,150)(50,-123,-55){3}{4}
\Text(68,73)[]{$1$}
\Text(155,73)[]{$2$}
\Text(81,118)[]{$1$}
\Text(142,118)[]{$2$}
\Text(110,15)[]{$(c)$}
%%%%%% (d) %%%%%%%%%%
\ArrowArcn(340,80)(40,180,0)
\ArrowArcn(340,80)(40,0,45)
\ArrowArcn(340,80)(40,45,135)
\ArrowArcn(340,80)(40,135,180)
\Photon(240,80)(300,80){3}{4}
\Photon(440,80)(380,80){3}{4}
%\Gluon(340,120)(340,40){3}{7}
\GlueArc(340,10)(50,55,123){3}{4}
\Text(297,88)[]{$1$}
\Text(386,91)[]{$2$}
\Text(311,42)[]{$1$}
\Text(370,42)[]{$2$}
\Text(340,15)[]{$(d)$}

\end{picture}\\
{Figure 2: {\sl Two loop photon polarization graph for the computation 
of $\gamma^*\rightarrow q\bar{q}g$ and $\gamma^*g\rightarrow
q\bar{q}$ amplitude .}}  

\end{center}

%%%%%%%%%%%%%% End Fig.2 %%%%%%%%%%%%%%%%

\newpage

%%%%%%%%% Fig.3 %%%%%%%%%%%%%%

\begin{center}
\begin{picture}(450,130)(0,0)
%\Line(0,0)(450,0)
%\Line(0,0)(0,130)
%%%%%% (a) %%%%%%%
\ArrowArcn(110,80)(40,180,90)
\ArrowArcn(110,80)(40,90,0)
\ArrowArcn(110,80)(40,0,90)
\ArrowArcn(110,80)(40,90,180)
\Photon(10,80)(70,80){3}{4}
\Photon(210,80)(150,80){3}{4}
\Gluon(110,120)(110,40){3}{7}
\Text(68,73)[]{$1$}
\Text(155,74)[]{$2$}
\Text(110,127)[]{$1$}
\Text(110,33)[]{$1$}
\Text(110,15)[]{$(a)$}
%%%%%% (b) %%%%%%%
\ArrowArcn(340,80)(40,180,90)
\ArrowArcn(340,80)(40,90,0)
\ArrowArcn(340,80)(40,0,90)
\ArrowArcn(340,80)(40,90,180)
\Photon(240,80)(300,80){3}{4}
\Photon(440,80)(380,80){3}{4}
\Gluon(340,120)(340,40){3}{7}
\Text(298,73)[]{$1$}
\Text(386,74)[]{$2$}
\Text(340,127)[]{$2$}
\Text(340,33)[]{$2$}
\Text(340,15)[]{$(b)$}

\end{picture}\\
{Figure 3: {\sl Two loop photon polarization graph for the computation 
of vertex correction of $\gamma^*\rightarrow q\bar{q}$ amplitude 
to $\alpha_s$ order.}}  

\end{center}

%%%%%%%%%%%%%% End Fig.3 %%%%%%%%%%%%%%%%5

\vspace{2.5cm}

%%%%%%%%%% Fig.4 %%%%%%%%%%%%

\begin{center}
\begin{picture}(450,250)(0,0)
%\Line(0,0)(450,0)
%\Line(0,0)(0,250)
%%%%%% (a) %%%%%%%
\ArrowArcn(110,200)(40,180,135)
\ArrowArcn(110,200)(40,135,45)
\ArrowArcn(110,200)(40,45,0)
\ArrowArcn(110,200)(40,0,180)
\Photon(10,200)(70,200){3}{4}
\Photon(210,200)(150,200){3}{4}
%\Gluon(110,240)(110,160){3}{7}
\GlueArc(110,270)(50,-123,-55){3}{4}
\Text(68,193)[]{$1$}
\Text(155,194)[]{$2$}
\Text(81,238)[]{$1$}
\Text(142,238)[]{$1$}
\Text(110,135)[]{$(a)$}
%%%%%% (b) %%%%%%%
\ArrowArcn(340,200)(40,180,135)
\ArrowArcn(340,200)(40,135,45)
\ArrowArcn(340,200)(40,45,0)
\ArrowArcn(340,200)(40,0,180)
\Photon(240,200)(300,200){3}{4}
\Photon(440,200)(380,200){3}{4}
%\Gluon(340,240)(340,160){3}{7}
\GlueArc(340,270)(50,-123,-55){3}{4}
\Text(298,193)[]{$1$}
\Text(386,194)[]{$2$}
\Text(311,238)[]{$2$}
\Text(371,238)[]{$2$}
\Text(340,135)[]{$(b)$}
%%%%%% (c) %%%%%%%%%%
\ArrowArcn(110,80)(40,180,0)
\ArrowArcn(110,80)(40,0,45)
\ArrowArcn(110,80)(40,45,135)
\ArrowArcn(110,80)(40,135,180)
\Photon(10,80)(70,80){3}{4}
\Photon(210,80)(150,80){3}{4}
%\Gluon(110,120)(110,40){3}{7}
\GlueArc(110,10)(50,55,123){3}{4}
\Text(68,73)[]{$1$}
\Text(155,73)[]{$2$}
\Text(81,42)[]{$1$}
\Text(142,42)[]{$1$}
\Text(110,15)[]{$(c)$}
%%%%%% (d) %%%%%%%%%%
\ArrowArcn(340,80)(40,180,0)
\ArrowArcn(340,80)(40,0,45)
\ArrowArcn(340,80)(40,45,135)
\ArrowArcn(340,80)(40,135,180)
\Photon(240,80)(300,80){3}{4}
\Photon(440,80)(380,80){3}{4}
%\Gluon(340,120)(340,40){3}{7}
\GlueArc(340,10)(50,55,123){3}{4}
\Text(297,88)[]{$1$}
\Text(386,91)[]{$2$}
\Text(311,42)[]{$2$}
\Text(370,42)[]{$2$}
\Text(340,15)[]{$(d)$}

\end{picture}\\
{Figure 4: {\sl Two loop photon polarization graph for the computation 
of self energy corrections of $\gamma^*\rightarrow q\bar{q}$ 
amplitude to $\alpha_s$ order.}}  

\end{center}

%%%%%%%%% End Fig.4 %%%%%%%%%

%%%%%%%%% Fig.5 %%%%%%%%%%%%%

%\begin{figure}[hbt]
\begin{figure}
%\vspace{-3.3cm}
\centerline{\psfig{file=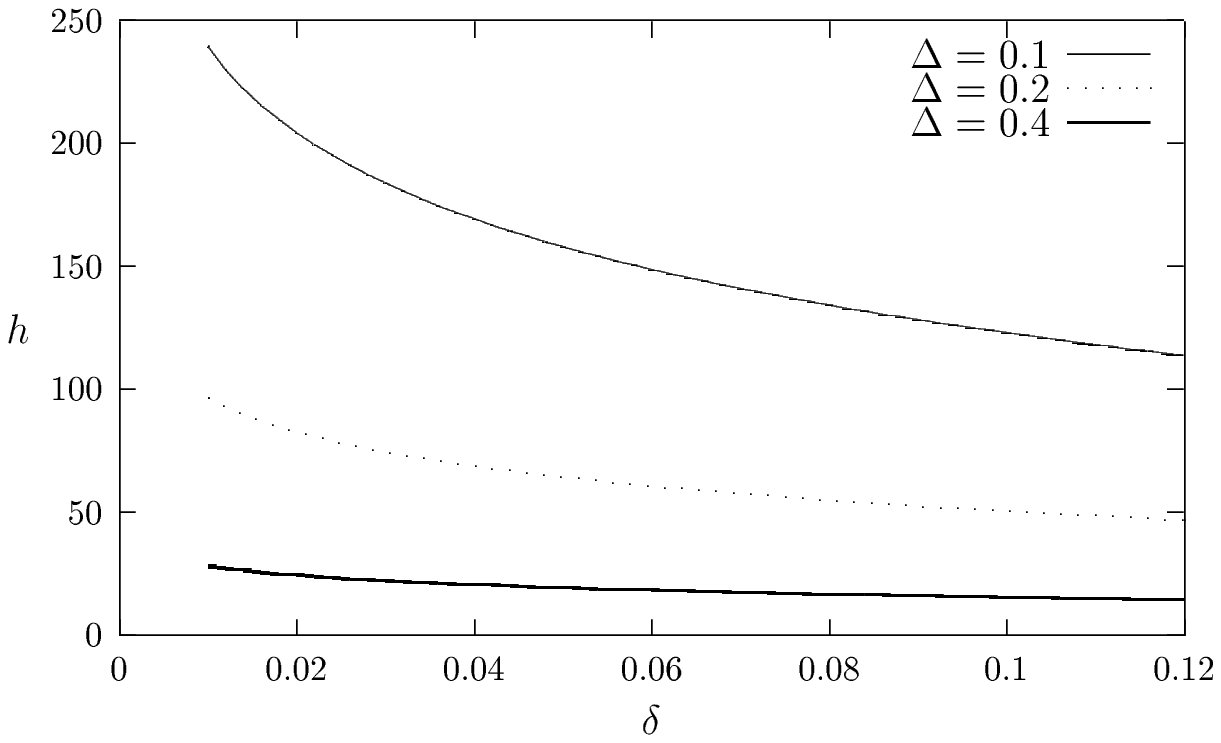,width=15cm}}
%\vspace{-12cm}
%%\caption{\sl{Plot of $h$ versus $\delta$ for different values of $\Delta$
%%and $\sqrt{q^2}/T=10$.}}
\begin{center}
{Figure 5: \sl{Plot of $h$ versus $\delta$ for different values of $\Delta$
and $\sqrt{q^2}/T=1.2$.}}
\end{center}
\end{figure}  

%%%%%%%%%% End Fig.5 %%%%%%%%%%

%%%%%%%% Fig.6 %%%%%%%%%%%%

%\begin{figure}[hbt]
\begin{figure}
%\vspace{-3.3cm}
\centerline{\psfig{file=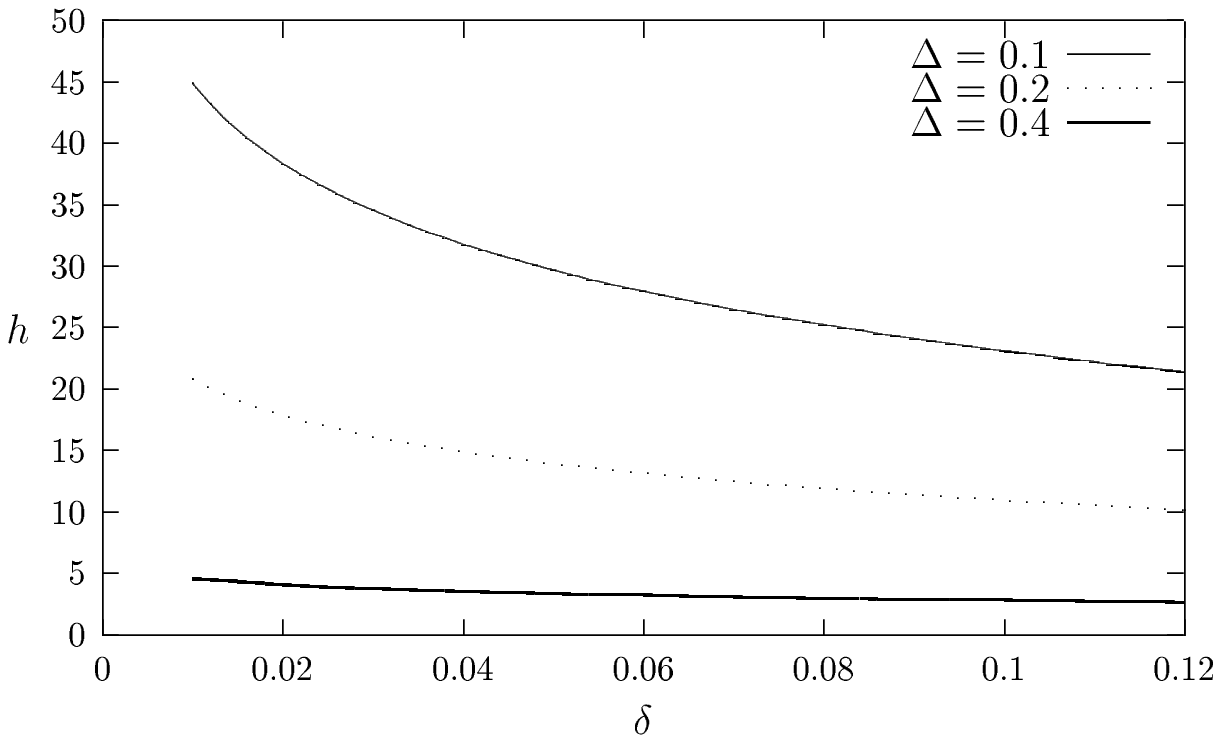,width=15cm}}
%\vspace{-12cm}
%%\caption{\sl{Plot of $h$ versus $\delta$ for different values of $\Delta$
%%and $\sqrt{q^2}/T=10$.}}
\begin{center}
{Figure 6: \sl{Plot of $h$ versus $\delta$ for different values of $\Delta$
and $\sqrt{q^2}/T=10$.}}
\end{center}
\end{figure}  

%%%%%%%% End Fig.6 %%%%%%%%%

\end{document}